\documentstyle[12pt,a4,epsfig]{article}

\newcommand{\beq}[1]{\begin{equation}\label{#1}}
\newcommand{\eeq}{\end{equation}}
\newcommand{\beqar}[1]{\begin{eqnarray}\label{#1}}
\newcommand{\eeqar}{\end{eqnarray}}
\newcommand{\r}{\vec{r}}

\newcommand{\al}{\alpha}

\newcommand{\ga}{\gamma}

\newcommand{\ka}{\vec{\kappa}}
\newcommand{\la}{\lambda}

\newcommand{\si}{\sigma}

\newcommand{\Ga}{\Gamma}

\begin{document}
\vspace*{-2cm}
\begin{flushright}
DESY 97--139\\
 NTZ 16/97\\
July 1997
\end{flushright}
\vspace{2cm}
\begin{center}
{\LARGE \bf 
Diffractive meson production from virtual photons
with odd charge-parity exchange  
\footnote{Supported by 
German Bundesministerium f\"ur Bildung, Wissenschaft, Forschung und
Technologie, grant No. 05 7LP91 P0, and in the framework of the German-Polish
agreement on scientific and technological cooperation, 
grant No. N - 115 - 95.}}\\[2mm]
\vspace{1cm}
R. Engel$^\S$, D.Yu.~Ivanov$^{\dagger \$}$, R.~Kirschner$^\dagger$ and  
L.~Szymanowski$^{\dagger \#}$\\
\vspace{1cm}
$^\S$ Deutsches Elektronen-Synchrotron DESY, D-22603 Hamburg, Germany
\\ 
\vspace{2em}
$^\dagger$Naturwissenschaftlich-Theoretisches Zentrum  \\
und Institut f\"ur Theoretische Physik, Universit\"at Leipzig
\\ 
Augustusplatz 10, D-04109 Leipzig, Germany
\\ 
\vspace{2em}
$^{\$}$
Institut of Mathematics, 630090 Novosibirsk, Russia \\
\vspace{2em}
$^{\#}$
Soltan Institut for Nuclear Studies, Ho\.za 69, 00-681 Warsaw, Poland
\end{center}

\vspace{1cm}
\noindent{\bf Abstract:}
We calculate the cross section of diffractive charge-parity
$C=+1$ neutral meson 
production in virtual photon proton collision at high energies. 
Due to the opposite
$C$-parities of photon and meson $M^+$ 
($M^+ = \eta_C, \pi^0, a_2$) this process 
probes the $t$-channel $C=-1$ odderon 
exchange which is described here as noninteracting three--gluon 
exchange. 
Estimates for the cross section of  
inelastic diffractive process $\gamma^* p\to \eta_C X_p$ are 
presented.
The total cross section of diffractive $\eta_C$ meson 
photoproduction is found to be $47$ pb. 
The cross sections for the diffractive production of
light mesons ($\pi^0, a_2$)
in $\gamma^* p$ collisions are of the same order if 
the photon virtuality $Q^2$ is  $ m^2_C$.

\vspace*{\fill}
\eject
\newpage

\section{Odd charge-parity exchange}

The increase of luminosity at HERA offers the possibility of
experimental investigations of diffractive processes of photo- and
electroproduction of neutral  charge-parity even
mesons $M^{+}=\pi^0, a_2, \eta_C$ at high energies
(see, for example, \cite{tapp}).
Studies of these  reactions
are of great interest since the opposite charge-parities $C$
of photon ($C=-1$) and 
 $M^{+}$ meson
cause definite $C$-parity  $C=-1$ of the  $t$- channel exchange.
Consequently the pomeron exchange being most important at high energies
does not contribute there. 
Therefore these processes allow for clean investigations 
of other Regge trajectories which 
are characterized by negative $C=-1$ parity. 

Reggeons having $C=-1$ are the $\omega$ trajectory  and the odderon. 
According
to fits to data on total hadron--hadron cross sections the intercept
of $\omega$ trajectory is close to 0.5 \cite{omega} and its contribution
decreases with energy. The odderon is the $C$ odd partner 
of pomeron with an intercept $\geq$ 1 which has been
introduced in phenomenology long time ago
\cite{odderon}. It is assumed that, like the pomeron,
 the odderon is related to gluonic
degrees of freedom. The odderon exchange results in 
a  difference between, for example, the total $pp$ and 
$p\bar p$ cross sections which does not decrease with increasing energy.   
Up to now there is no indication of that difference in present
high energy  $pp$ and $p\bar p$ data.  
Also odderon effects have not been observed in other soft hadronic 
collisions.
 
The reasons for the absence of the odderon exchange in soft hadronic reactions 
remains to be understood since in perturbative QCD the pomeron and the odderon 
appear on the equal footing.
In Leading Logarithmic Approximation (LLA) of 
perturbative QCD the pomeron corresponds 
to the exchange of two interacting reggeized
gluons (hard pomeron) whereas the odderon is 
described by exchange of three interacting 
reggeized gluons (hard odderon), see \cite{odd-theory}. 
In Born approximation, i.e.\ considering the pure three--gluon exchange  
in the $t$-channel, the intercept of the odderon is equal to 1.
The effect of gluon interactions is expected  to increase of 
odderon intercept \cite{GLN}.
Therefore the hard odderon should manifest itself in hard 
diffractive processes at high energies. 

The idea to study odderon exchange dominated processes
in the diffractive meson 
production was considered by Sch\"afer, Mankiewicz and Nachtmann in
Ref.~\cite{Nacht} where the odderon exchange was treated purely 
phenomenologically, in analogy to the one-photon exchange. 
The possibility to probe the hard odderon in processes of light meson 
production in photon-photon collisions at large $t$ 
was considered in \cite{GinIv}. 
Recently exclusive $\eta_C$
photoproduction on protons was 
studied by Czy\.zewski {\it et al.} in \cite{kwiec} 
describing the odderon as three-gluon exchange. 

In this paper we shall calculate the diffractive production of $C=+1$ mesons
in Born approximation, i.e. involving the exchange of three 
noninteracting gluons in the 
$t$-channel. 
We use assumptions similar to the ones made in \cite{kwiec}, though 
our method  is  different. 
It is based on the impact factors obtained in \cite{GinIv}. 
The results for heavy mesons 
are compared with \cite{kwiec} and discrepancies are discussed.
Furthermore we study  diffractive production of light
mesons  $M^{+}=\pi^0, a_2, \dots$ in virtual photon-proton interactions and 
the process of diffractive  meson production at large $t$
which is characterized by proton disintegration.

\section{Impact factors and meson wave functions}

In the following we consider diffractive meson production processes 
which can be characterized by at least one hard scale. 
This scale is either the  mass of a heavy quark
(for example, in $\eta_C$ production) or the virtuality $Q^2$ of the photon 
in the production of light mesons.

The quasi-elastic production of a $M^+$ meson is shown in Fig.~1.
The mean transverse distance of the quarks in the upper block of 
Fig.~1 is $\sim 1/m_c, 1/Q$. 
The  other essential (soft) scale in our problem  is the mean transverse 
separation between the quarks
in proton $\sim 1/m_{\rho}$ (lower part of Fig.~1). 
With the growth of the transfered momentum 
$|t|\geq m^2_{\rho}$ the process with photon disintegration
shown in Fig.~2 becomes more important
than  the elastic one (Fig.~1). In the region of 
large $t$ ($|t|\gg \Lambda^2_{\rm QCD}$) the cross
section for this process can be expressed in terms of the 
photon-quark cross section 
and the quark densities inside the proton. In this case there 
is no soft scale and we deal
with a purely hard diffractive process where the perturbative 
odderon should contribute.  
In contrast, the production of light mesons by real photons
is a typical soft process and is not considered here.

The amplitude determined by three--gluon 
exchange can be written in 
impact representation as an integral over the gluon transverse
momenta $\ka_{i}$ \cite{GinIv}  
\beq{impfac}
{\cal M} = \frac{s}{3(2\pi)^4}\int \frac{d^2\kappa_1 \; d^2\kappa_2}
{\vec{\kappa}^{\,2}_1\, \vec{\kappa}^{\,2}_2\,\vec{\kappa}^{\,2}_3}
J_{\gamma^* M}(\vec{\kappa}_1,\vec{\kappa}_2; \vec{q})\cdot
J_{pp}(\vec{\kappa}_1,\vec{\kappa}_2; \vec{q}) \ .  
\eeq
The kinematical notations are given in Fig.~1.
Here $s=(q+p)^2$ and $t=(p_1-p_M)^2 =  - \vec{q}^{\,2}$ is 
the momentum transfer with
$ \vec{q}$ = $\ka_{1}+\ka_{2}+\ka_{3}$. 
The impact factors $J_{\gamma M}$ and $J_{pp}$ 
correspond to the upper and lower part of Fig.~1, correspondingly.

The impact factor $J_{\gamma^* M}$ for the $\ga^* \to M$ transition 
can be derived within perturbative QCD. It can be expressed 
in terms of the amplitude $J_{\gamma^*
 q\bar{q}}$ for $q\bar{q}$ production which
has the form \cite{GinIv} 
\beqar{quark-im}
J_{\gamma^* q\bar{q}} &=& eg^3Q_q \left(\frac{d^{abc}}{4N}\right)
\bar{u}_1 \left[ \; \cdots \;\right] \frac{\hat{p}_2}{s} u_2     \nonumber \\
\left[ \;\cdots \;\right] &=& m R \hat{e} - 2x(\vec{Q}\vec{e}) -
\hat{Q}\hat{e} \nonumber \\
R &=& \left( \frac{1}{m^2_0 + \vec{q}^{\,2}_1} - \sum^3_{i=1}
\frac{1}{m^2_0 + (\vec{\kappa}_i - {\vec{q}_1)}^{2}} \right) - \left(\vec{q}_1
\leftrightarrow \vec{q_2} \right) \nonumber \\
\vec{Q} &=& \left( \frac{\vec{q}_1}{m^2_0 + {\vec{q}_1}^{\,2}} + \sum^3_{i=1}
\frac{\vec{\kappa}_i - \vec{q}_1}{m^2_0 +
(\vec{\kappa}_i - \vec{q}_1)^2} \right) + \left(\vec{q}_1
\leftrightarrow \vec{q_2} \right) \nonumber \\
m^2_0 &=& m^2 + Q^2x(1-x)
\eeqar
In the above expression
$e^\mu=(0,\vec{e},0)$ is the polarization vector of the photon.
Furthermore we use the notations
 $\al =e^2/4\pi = 1/137$ and  $\al_s =g^2/4\pi$. $Q_qe$ is the 
quark charge and $d^{abc}$ are the totally symmetric structure constants for 
the $SU(N)$ color group. 
$\vec q_1$ and $\vec q_2$ are the transverse momenta of
quark and antiquark, $0<x<1$ is the fraction of the 
quark longitudinal momentum.
The virtuality of photon is $p_1^2=-Q^2$, $m$ is the quark mass.
The quark spinors are denoted by $u_1=u(q_1)$, $u_2=u(q_2)$.
It is convenient to introduce the variables
\beq{xi}
\xi = 2x -1 \;\;,\;\;\;\nu = \frac{4m^2+Q^2(1-\xi^2)}{\vec q^2} \ .
\eeq

We will neglect the transverse motion of quarks inside the meson (collinear
approximation). Then the $q\bar q\to M$ transition is described by  the 
Eq.~(\ref{quark-im}) in which the 
quark spinors are substituted by the meson wave 
function $\varphi_M(\xi )$ (see e.g. \cite{ChZi,BaGr}).
One obtains the substitutions:

(a) for the tensor meson $T$ with helicity $\lambda =0$
\begin{equation}
Q_q\bar u_1\dots u_2\to \frac{Q_T}{4N}\int \limits_{-1}^{+1}d\xi
f_T \varphi _T(\xi )  Tr\left( \dots\hat p_3\right)
\label{8}
\end{equation}

(b) for the pseudoscalar mesons $P$
\begin{equation}
Q_q\bar  u_1\dots u_2\to \frac{Q_P}{4N} \int\limits_{-1}^{+1}d\xi
f_P\varphi _P(\xi )Tr\left( \dots \gamma ^5\hat p_3\right). \label{9}
\end{equation}

The trace is performed  over  vector  and  color  indices. 
The quantity $Q_M$ ($M=P,T$) is the average quark charge in the meson $M$.
For instance
$|\pi^0 > =(|u\bar u > - |d\bar d >)/\sqrt{2}$, therefore
$$Q_\pi=
{1 \over \sqrt{2} }(+\frac{2}{3}-(-\frac{1}{3}))={1 \over \sqrt{2} } \ .$$
It should be noticed that the
production of tensor  mesons with the helicity 
$\lambda =\pm 1,2$ is suppressed in LLA at least by a factor 
$\sim \left(\Lambda_{\rm QCD}^2/Q^2\right)$ \cite{GinIv}.

According to Ref.\cite{ChZi} we adopt for numerical calculations  the
following parametrizations for the wave functions and
the coupling constants:

(a) for light tensor mesons $T$
\begin{eqnarray}
& \varphi _T(\xi )=\frac{15}{4}\xi \left( 1-\xi ^2\right)\,;
\quad f_T=85\!\mbox{ MeV }\,; & \label{10} \\
& Q_{a_2}=\frac{1}{\sqrt{2}}\ ,\quad Q_{f_2}=\frac{1}{3\sqrt{2}}\ ,\quad
Q_{f'_{2}}=\frac{1}{3} &\nonumber
\end{eqnarray}
where an ideal mixing is assumed, i.e. $f'$  consists of 
$s\bar s$-states only.

(b) for light pseudoscalar mesons $P$
\begin{eqnarray}
& \varphi _P(\xi )  = 
  \left\{
\begin{array}{ll}
\frac{3}{4} (1-\xi^2) & \mbox{asymptotical form} \\ 
  & \\                 
\frac{15}{4}\xi^2(1-\xi^2) & \mbox{Chernyak-Zhitnitsky form}
\end{array}
\right. 
\, ;  & \label{10'}\\
& \quad f_{\pi }=133\!\mbox{ MeV },\quad f_{\eta }  =  150\!\mbox{ MeV },\quad
f_{\eta '}=110\!\mbox{ MeV }\,;  & \label{11} \\
& Q_{\pi }=\frac{1}{\sqrt{2}}\ ,\quad Q_{\eta}  =  0.38\ ,\quad
Q_{\eta '}=0.14 &  \nonumber
\end{eqnarray}
where the standard $\eta\eta '$- mixing is taken into account.

(c) for the $\eta_C$ meson
\beq{eta}
 \varphi_{\eta_C} = \delta(\xi) \ , f_{\eta_C}=400 \mbox{ MeV }. 
\eeq
The photon width of $\eta_C$ is expressed through $f_{\eta_C}$ as
$\Ga _{\eta_C\to \ga\ga}= 
e_c^4f_{\eta_C}^2/(4\pi m_{\eta_C})\approx7\mbox{ keV }$ \cite{width}. 

The usage of a delta function for the wave function of heavy mesons is 
equivalent to the non-relativistic approximation. In the case of light 
meson production we take into account the nontrivial longitudinal motion
of the light quarks.

The conditions of the collinearity of quark momenta are
\begin{equation}
\vec q_1=\frac{1}{2}(1+\xi )\vec q,\quad
\vec q_2=\frac{1}{2}(1-\xi )\vec q,\quad
 \label{18}
\end{equation}
With these  relations  we  
insert Eqs.~(\ref{8},\ref{9}) into
Eq.~(\ref{quark-im}) and obtain the
impact-factors $J_{\gamma T},\, J_{\gamma P}$
\begin{equation}
J_{\gamma^* T}\left(\ka_{1,2},\vec q\right)=eg^3\left(
\frac{d^{abc}}{4N}\right)\frac{Q_T}{2}\int
\limits^{+1}_{-1}d\xi 
f_T\varphi _T(\xi )\xi (\vec Q\cdot\vec e)  \label{19}
\end{equation}
\begin{equation}
J_{\gamma^* P}\left(\ka_{1,2},\vec q
\right)=eg^3\left(
\frac{d^{abc}}{4N}\right)\frac{Q_P}{2}\int
\limits^{+1}_{-1}d\xi
f_P\varphi _P(\xi )\left[\vec e\times\vec Q\right] \label{20}
\end{equation}

Introducing the dimensionless  vectors $\vec r_i$
\begin{equation}
\ka_{i}=\frac{1}{2}\left(\vec r_i+\vec n\right){\mid \vec q \mid}\, ,
\quad \vec n=\vec q/{\mid \vec q\mid} \label{21}
\end{equation}
the
impact-factors (\ref{19},\ref{20}) read
\begin{equation}
J_{\gamma^* M}\left(\ka_{1,2 },\vec q\right)=eg^3\left(
\frac{d^{abc}}{4N}\right)Q_M\frac{f_M}{\mid q\mid}\left\{
\begin{array}{lll}
\left(\vec e\cdot\vec F_T\right) & \mbox{ for } & M=T\\
\left[\vec e\times\vec F_P\right] & \mbox{ for } & M=P
\end{array}\right. \label{22}
\end{equation}
\beqar{Fu}
\vec F_M&=&\int \limits^{+1}_{-1}d\xi \left\{\left[\frac{\vec n(1+\xi)}
{\nu + (1+\xi )^2 }+
\sum^3_{i=1}\frac{\vec r_i-\vec n\xi}{\nu + \left(\vec r_i-\vec n\xi\right)^2}
\right]+\left[\xi \leftrightarrow -\xi\right]\right\} \nonumber \\
&\cdot& \left\{
\begin{array}{lll}
\xi\varphi _T(\xi ) & \mbox{ for } & M=T \\
\varphi _P(\xi ) & \mbox{ for } & M=P
\end{array}\right. \nonumber
\eeqar
It should be noted that $\vec F_M\left(\vec r_i,\vec n\right)\to 0 $
at $\vec r_i\to -\vec n$.

The impact factor $J_{pp}$ which describes the lower 
part shown in Fig.~1 cannot be 
calculated in the region of small momentum transfers ($|t|\leq
\Lambda^2_{\rm QCD}$) within perturbative QCD since  the mean transverse
distance  between the quarks inside the  proton is large. Therefore, 
a phenomenological ansatz is used for this quantity. 
There are several restrictions
on this impact factor imposed by gauge invariance: $J_{pp}$ must vanish
when $\ka_{i}$ ($i=1,2,3$) tends to zero. This property ensures the
convergence of the integral in Eq.~(\ref{impfac}). 
Bose symmetry demands that 
$J_{pp}$ should be symmetric under interchange of $t$-channel gluons.  
Of course, both impact factors $J_{\gamma q\bar{q}}$ and $J_{\gamma
M}$ discussed above satisfy these requirements.   

For $J_{pp}$  we use the expression
derived by Fukugita and Kwieci\'nski within the eikonal
model \cite{FuKw}. In our conventions it reads
\beqar{FuKw}
&& J_{pp}(\ka_1,\ka_2,\ka_3) = \nonumber \\
&& \bar{g}^3 \left(\frac{d^{abc}}{4N}\right)\cdot 3 \cdot
\left[ F(\vec{q},0,0) - \sum_{i=1}^3 F(\ka_i, \vec{q} -\ka_i,0) +
2F(\ka_1,\ka_2,\ka_3)\right]
\eeqar
where $\vec{q} =\ka_1 +\ka_2 +\ka_3$ and
\beq{F}
F(\ka_1,\ka_2,\ka_3) = \frac{A^2}{A^2 + \frac{1}{2}[(\ka_1 - \ka_2)^2 + (\ka_2 -
\ka_3)^2 + (\ka_3 - \ka_1)^2]}
\eeq
with $\bar{g}^2/(4\pi) \approx 1$ and $A = m_\rho/2$.
This eikonal model describes correctly the values of total hadronic 
cross sections \cite{GuSo,FuKw}. The normalization of Eq.~(\ref{FuKw}) can be 
understood in the following way. The three-gluon impact factor of point 
like fermions does not depend on the gluon momenta. Hence this 
coupling is given by \cite{GinIv}
\beq{quark}
J^{3G}_{qq}=g^3\left(\frac{d^{abc}}{4N}\right) \ .
\eeq
The first term in brackets in Eq.~(\ref{FuKw}) results from 
diagrams where all the three $t$-channel gluons couple to the same quark line.
When $q$ is equal to zero the interaction with the $t$-channel gluons does not
disturb quark motion and, due to the normalization of proton wave function,
this part of the proton impact factor is equal to
$3\cdot g^3\left(\frac{d^{abc}}{4N}\right)$ where the factor 3  
counts the number of constituent  quarks inside the proton. 
In the case of the proton electromagnetic form
factor this coefficient 3 will be changed to 1 due to the fractional
electric charges of quarks $1=2/3+2/3-1/3$ and the gauge group factor is 
replaced by 1. The other terms in 
brackets in Eq.~(\ref{FuKw}) describe the contribution of the other diagrams
where gluons couple to different quark lines. The whole expression 
(\ref{FuKw}) is gauge invariant and obeys Bose symmetry.
In terms of dimensionless variables the proton impact factor reads
\beq{prim}
J_{pp}=3\bar g^3\left(\frac{d^{abc}}{4N}\right)F_p
\eeq
\beqar{dimen}
 F_p &=&\left[
\frac{1}{1+4z}- \sum_{i=1}^{3}\frac{1}{1+z(3\r^2_i+1)}+ \right. \nonumber \\ 
& &\left. \frac{2}
{1+\frac{z}{2}[(\r_1-\r_2)^2+(\r_2-\r_3)^2+(\r_3-\r_1)^2]}  
\right]
\eeqar
with $z=\frac{\vec q^2}{m^2_{\rho}}$, $\r_3=-\r_1-\r_2-\vec n$.

Substituting the impact factors (\ref{22}, \ref{prim}) into 
Eq.~(\ref{impfac}) the following
expression for the amplitude is obtained
\beq{final}
{\cal M}_{\gamma^* p \to M p} = eg^3(\mu^2 )
\bar g^3Q_M\frac{5}{18\pi^2}\frac{sf_M}
{|t|^{3/2}}I  \cdot 
\left\{
\begin{array}{ccc}
[\vec e \times \vec n] & for & M=P \\
(\vec e \cdot \vec n) & for & M=T
\end{array}
\right.
\eeq
where
\beq{I}
I=\frac{1}{12\pi^2}\int \frac{d^2r_1d^2r_2}
{(\r_1+\vec n)^2(\r_2+\vec n)^2(\r_1+\r_2)^2}F_M F_p
\eeq

In the case of $\eta_C$ production Eq.~(\ref{final}) can be compared with 
the results reported in \cite{kwiec}. Our amplitude (\ref{final}) is 
3 times smaller. We have given here our arguments about the normalizing 
coefficients in all details in order to show that our normalization is 
the correct one.  

As already mentioned in the introduction, at large $t$ 
the diffractive process 
with proton dissociation $\gamma p\to MX $
dominates over the elastic one. In this case 
the cross section can be expressed in terms of 
 the cross section of photon-quark
scattering and the quark densities inside proton (Fig.~2)     
\beq{disint}
\frac{d\si}{dt\; dx} = \sum_{q_i ,\vec{q}_i} 
\frac{d\hat{\si}_{\ga q \to M q}}{dt} \left[ q_i(x,t) + \bar{q}_i(x,t)
\right]
\eeq
It is important to notice that the three-gluon $C$-odd 
odderon does not couple to gluons in the proton \cite{Gin}.

\section{Results on diffractive meson production}

The results which are presented below have been obtained by numerical 
evaluation of the integral $I$ in Eq.~(\ref{final}) and of 
 a similar integral for hard diffractive scattering on quarks. 
The hard cross section 
$d\hat{\si}_{\ga q \to M q}/dt$ was calculated from the amplitude 
(20) with the replacement $g^3\bar g^3\to g^6(\mu^2)$, 
where $\mu$ is the scale for the evaluation of the running
strong coupling constant. 

Let us start with the
discussion  of the results for elastic
$\eta_C$ production.
In the calculations we use the scale
$\mu^2=m_c^2$, $m_c=1.4 \mbox{ GeV }$ and $\Lambda^{(4)}_{\rm QCD}
=0.2\mbox{ GeV }$ (for 4 flavors).
Our results for the unpolarized differential cross section for the process
$\gamma^* p\to \eta_C p$ are given in Fig.~3. This cross section
vanishes at $t=0$, 
it reaches the maximum at $|t|\approx 0.5 \mbox{ GeV }$ and decreases 
 for larger $|t|$. The differential cross section 
decreases rapidly with the growth of photon virtuality.      
In Fig.~3  we show also the result for $Q^2=0$ 
reported in Ref.~\cite{kwiec} which 
is approximately 4 times larger than ours. 
The reasons for this discrepancy are the following:
(i) as discussed above in detail, the amplitude (20) turns out to be 
smaller in normalization by a factor of 3 compared to the result 
quoted in \cite{kwiec}, and (ii)
we use $\alpha_s(m_c^2) \approx 0.387$ instead of 0.3 as done in
\cite{kwiec}.

In Fig.~4 we present the cross section integrated over the region 
$0<|t|<t_0$ as a function of $t_0$. We see that the main part of 
total cross section originates from the region of small $|t|$. 
For real photoproduction the total cross section is about $47$ pb,
it decreases with increasing $Q^2$.

Now let us discuss the results for large momentum transfer $t$ 
when the proton disintegrates.
In this case the coupling constant has been
 evaluated at the scale $\mu^2~=~|t|$.
The cross section of photon-quark scattering 
was convoluted with the GRV parametrization \cite{GRV94}
of the quark densities in the proton, see Eqs.~(\ref{disint}). 
The results are given in Figs.~5 and 6. Comparing these figures
with the corresponding cross sections for elastic $\eta_C$ production 
(Figs.~3, 4)
one can see that the process $\gamma^* p\to \eta_C X_p$ is not  
considerably suppressed. For example, its cross section is 11~pb
for $|t|>3 \mbox{ GeV}^2$ and $x\geq 0.1$.        

Calculating light meson production we have an additional nontrivial 
integration over the variable $\xi$, 
describing the longitudinal motion of light
quarks.
The results for the cross sections of $\pi^0$ and  $a_2$ meson
production are shown in Figs.~7 and 8.
The cross sections for the production of other mesons can be readily 
obtained by changing corresponding coupling constants given in 
Eqs.~(\ref{10},\ref{11}). 
The strong coupling constant has been evaluated at the scale $\mu^2=Q^2+|t|$.
In the case of $\pi^0$ production the asymptotic form of the  wave
function (\ref{10'}) has been used. Comparing Eqs.~(\ref{22}) 
and (\ref{9},\ref{10})
it can be seen that the formula for the amplitude of 
tensor meson production
is similar to one for the production of pseudoscalar mesons described 
by the  Chernyak-Zhitnitsky wave function. They differ only
in the meson coupling 
constants and in the polarization dependence. The unpolarized cross sections  
differ by the coupling constants only, e.g. 
$d\sigma_\pi=(f_\pi/f_{a_2})^2d\sigma_{a2}$.         
Comparing the Figs.\ for $\pi^0$ and $a_2$ production 
it can be seen that the shape of differential 
cross sections depends strongly 
on the form  of the meson wave function. The position of the  dip is very 
sensitive to the form of the wave function of the light meson. It gets
shifted to larger $|t|$ with increasing photon virtuality $Q^2$.   

In Tab.~\ref{table} the cross sections of diffractive meson production
are compared.
The cross sections of $\pi^0$ and $ a_2$ meson production are comparable 
to the cross section of $\eta_C$ production for $Q^2\approx m_c^2$.
They fall off very rapidly with increasing $Q^2$.
\begin{table}[htb]
\caption{\label{table} 
\it
Table of cross sections for diffractive meson
production (for light mesons $|t| > t_0$ = 3 GeV$^2$). The cross
sections are given in pb.}
\medskip
\renewcommand{\arraystretch}{1.3}
\begin{tabular}{||c|c|r|c|r|c||}\hline\hline
meson / $Q^2$ [GeV$^2$] &  0   &   1  &   2  &    5  &   10  
\\ \hline\hline
$\eta_C$                &  47  &  33  &  24  &   11  &   4.2
\\ \hline
$\pi^0$                 &  -   &  -   & 100  &   8.4 &   1.1
\\ \hline
$a_2$                   &  -   &  -   &  31  &   3.6 &   0.68
\\ \hline\hline
\end{tabular}
\end{table}

Performing the calculations for the
diffractive production of light mesons
we have to take care careful about the lower  
limit of the photon virtuality $Q^2$ which to ensure the 
self-consistency of our perturbative approach.  To get an idea 
about this limit the numerical calculations of the cross 
sections have been done keeping in our formulae a non zero quark 
mass $m$ which can be  
considered as the  measure of non-perturbative effects. We have chosen  
two typical values of the constituent quark masses
$m=0.2 \mbox{ GeV}$ and $m=0.3 \mbox{ GeV}$.  
In the Table \ref{table2} the results for the ratio 
of the total cross sections $\sigma(m^2)/\sigma(m^2=0)$
for tensor meson production are given
for some values of photon virtualities $Q^2$.
\begin{table}[htb]
\caption{\label{table2}
\it Ratio 
of the total cross sections $\sigma(m^2)/\sigma(m^2=0)$
for elastic tensor meson production for different photon virtualities.
}
\medskip
\renewcommand{\arraystretch}{1.3}
\begin{tabular}{||c|c|c|c|c||}\hline\hline
quark mass / $Q^2$ [GeV$^2$] &  1   &   2  &   5  &  10
\\ \hline\hline
$m=200 \mbox{ MeV}$          & 0.69  & 0.90 &  0.98 &   0.99     
\\ \hline
$m=300 \mbox{ MeV}$          & 0.50  & 0.74  & 0.92  &  0.95
\\ \hline\hline
\end{tabular}
\end{table}
It can be concluded that perturbative calculations should be reliable
for $Q^2 \stackrel{>}{\sim} 2 \mbox{GeV}^2$.

\section{Discussion}

Summarizing we would like to stress that the numerical results presented
here are based on amplitudes the essential part of which have been
derived from perturbative QCD. In particular these are the impact 
representation (1) and the virtual photon impact factor (2). As further 
inputs we need the meson wave functions and couplings which are well known 
from phenomenology. 

The proton impact factor in the quasi-elastic diffraction which we have
adopted following \cite{FuKw} can be motivated by perturbative
calculations. However there is no hard scale in that part of the process
and the form of the proton impact factor enters as an additional
assumption.
The form of $J_{pp}$ (proposed by Fukugita and Kwiecinski 
\cite{FuKw})
exhibits the following features: it is gauge invariant, Bose symmetric
and incorporates the parameter $\bar g$ (frozen strong coupling) and 
a parameter related to the size of the proton $A \approx m_\rho/2$.
We suppose that the uncertainty related to its  
form does not affect seriously our  
value of the differential cross section in the region of small
momentum transfers and also not the value of total cross section since it 
is saturated mostly by the region of small $t$ (see Figs.~3, 4).
But it might be that at large momentum
transfers $|t|\gg\Lambda^2_{\rm QCD}$ 
this impact factor has to be changed since it does not obey the dimensional 
counting rules.

In inelastic diffraction being characterized by proton disintegration
the quark
distributions in the proton take the role played by the proton impact
factor before. This non-perturbative input is, of course, much better known.
Due to that this process has less theoretical uncertainties,
but  its cross section 
in the region $t_0<|t|, t_0=2\div 3 \mbox{ GeV}$ is smaller 
than the total cross section for the elastic process. 
 
Our calculations are done in the Born approximation. One can try to estimate 
roughly the effects related to the  
interaction of $t$-channel gluons by multiplying the formulae for
cross sections  by the
Regge factor $(s/\Lambda^2)^{2(\la_{odd}-1)}$ .
Using a value for the odderon intercept of $\la_{odd}=1.1$ and  
$\Lambda^2 \sim 
m_{\eta_c}^2, Q^2$  this factor can be as large as 5 at HERA energies.  

The value of the total $\eta_C$ photoproduction cross section in Born
approximation is $47$ pb. Similar cross sections
are obtained for the production of light mesons by virtual photons, if the
photon virtuality $Q^2$ is of the order $m_c^2$. 
It might be difficult to observe reactions with such cross sections at HERA. 
On the other hand the situation is much better if the odderon 
intercept is greater than 1 resulting in cross sections being 
a few times larger. 
Therefore we expect that experimental searches 
for diffractive production of $C=+1$ mesons  
 at HERA can bring new interesting results relevant for the longstanding
and controversial questions about the odderon.

\vspace*{1cm}
{\Large \bf  Acknowledgments}

\vspace*{.8cm}

D.Yu I. and L.Sz. would like to acknowledge the warm hospitality 
extended to them at University of Leipzig. 
One of the authors (R.E.) was supported in parts by the Deutsche
Forschungsgemeinschaft under contract No Schi 422/1-2.


\clearpage
\section*{Figures}


\hspace*{3cm}

\begin{figure}[htb]
\begin{center}
\epsfig{file=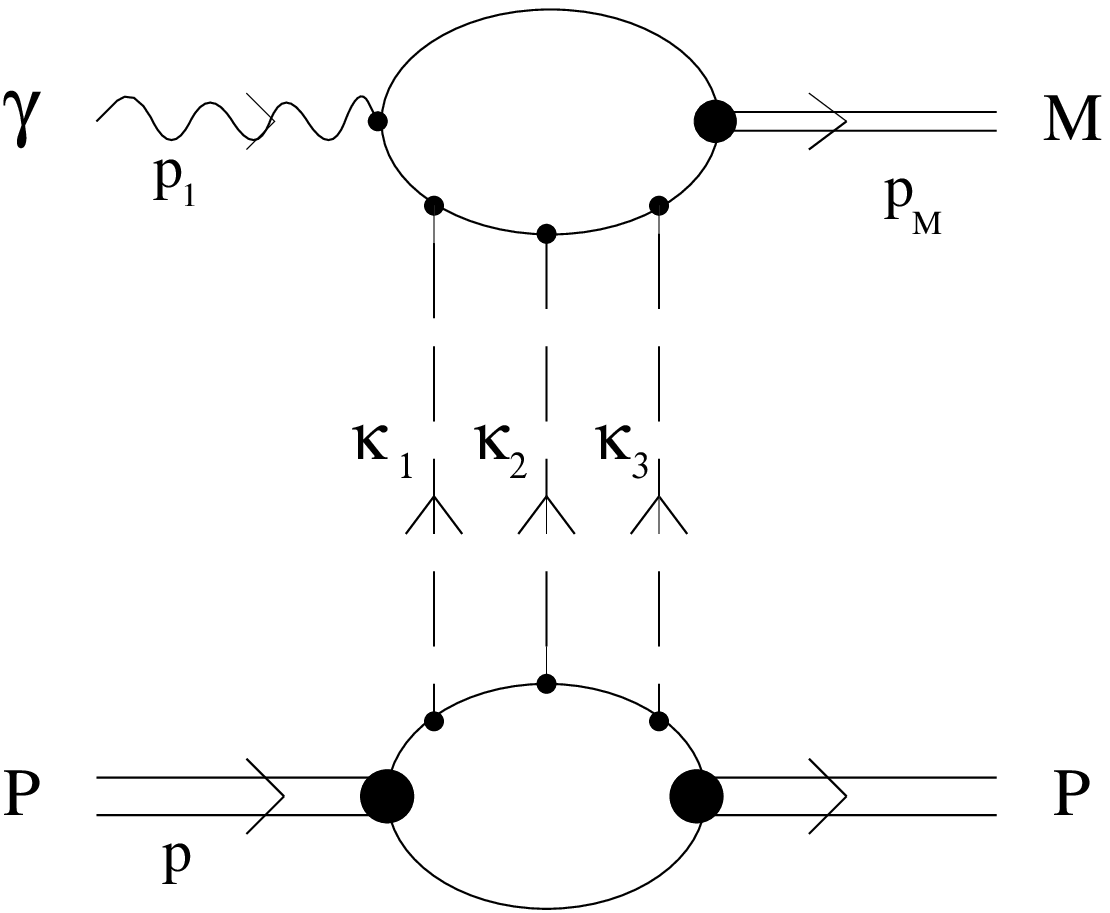,width=9cm}
\end{center}
\vspace*{0.5cm}
\caption{\label{fig-1}
The kinematics of elastic diffractive process $\gamma^* p \to M p$.
}
\end{figure}


\begin{figure}[htb]
\begin{center}
\epsfig{file=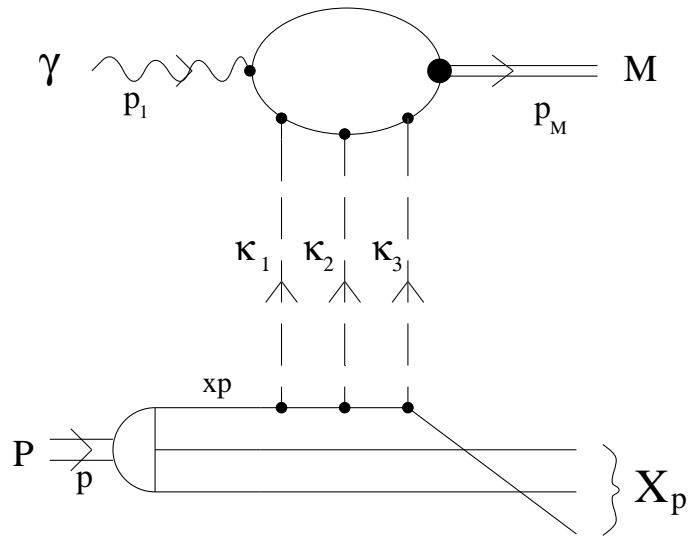,width=9cm}
\end{center}
\vspace*{0.5cm}
\caption{\label{fig-2}
The hard  diffractive process $\gamma^* p \to M X_p$.
$X_p$ are the hadrons originating from  proton disintegration.
}
\end{figure}


\begin{figure}[htb]
\begin{center}
\epsfig{file=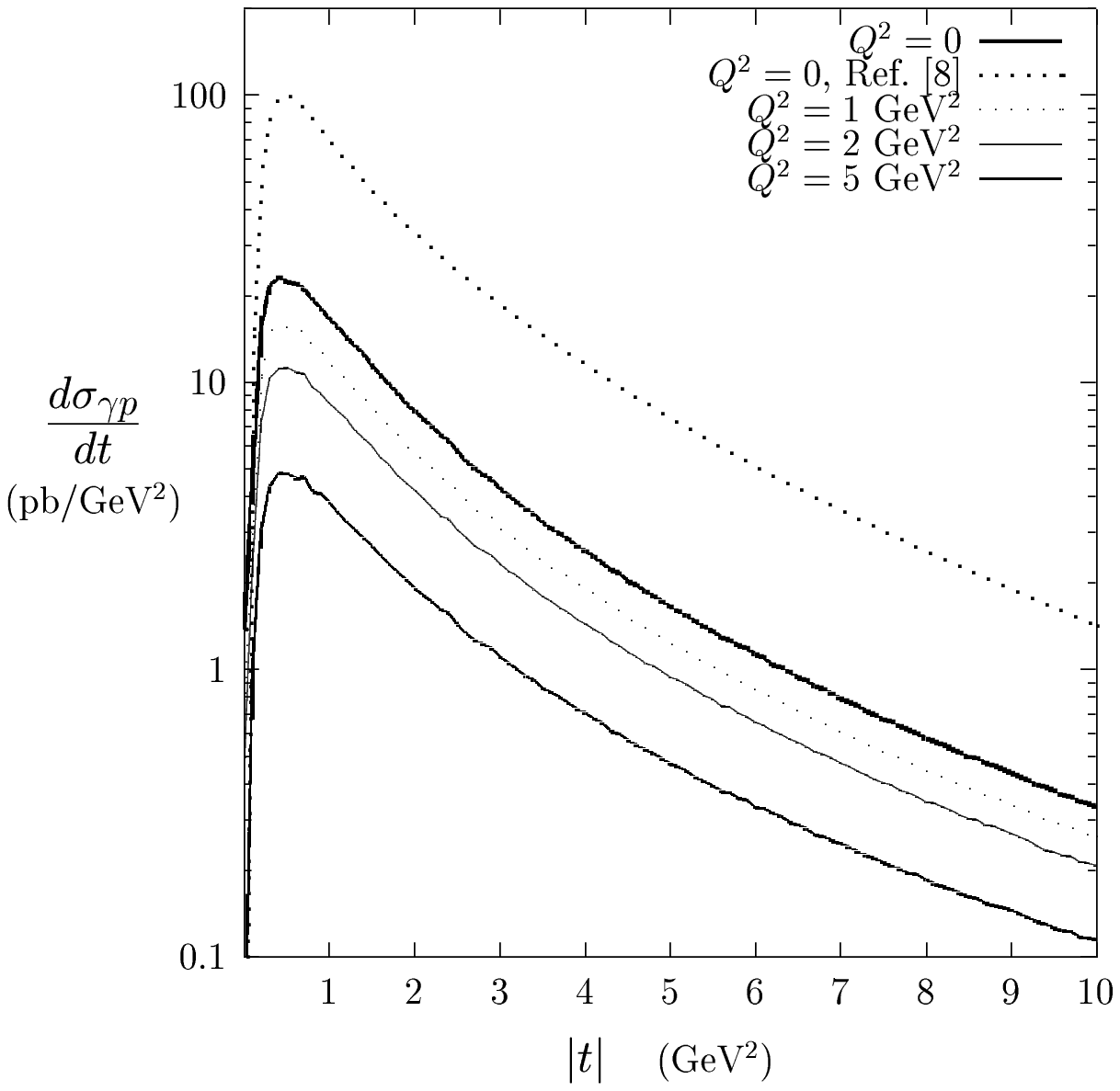,width=12cm}
\end{center}
\vspace*{0.5cm}
\caption{\label{fig-3}
Differential cross section for $\gamma^* p \to \eta_C p$. The strong 
coupling has been evaluated with the scale $\mu^2=m_c^2$ and with 
$\Lambda^{(4)}_{\rm QCD}=0.2 \mbox{ GeV}$.
}
\end{figure}


\begin{figure}[htb]
\begin{center}
\epsfig{file=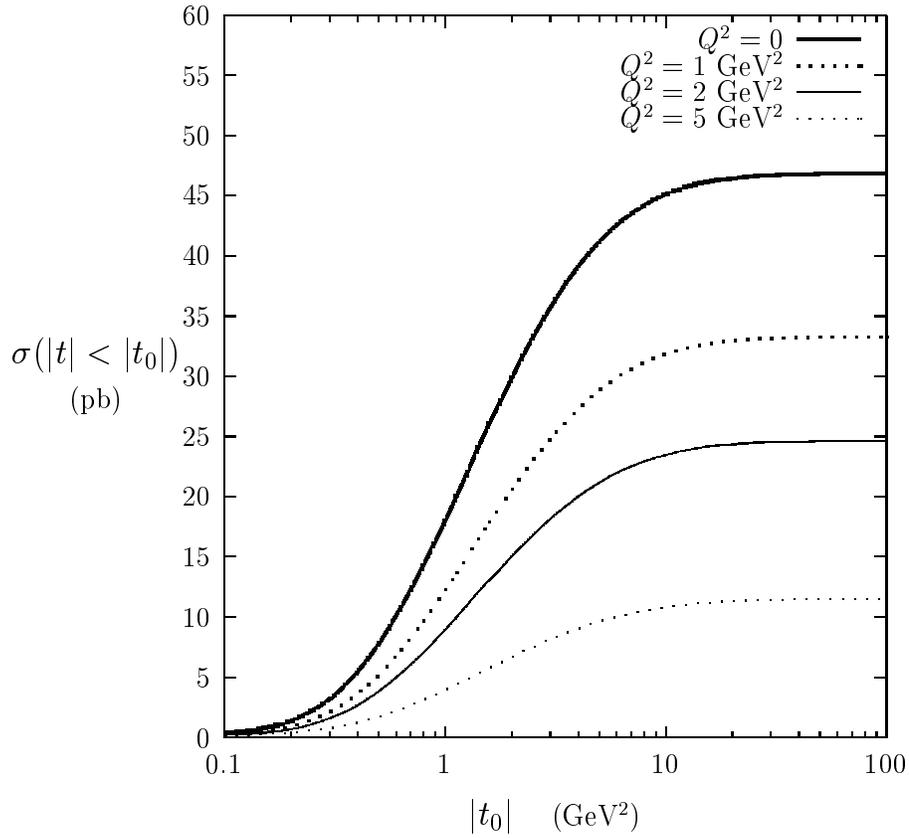,width=12cm}
\end{center}
\vspace*{0.5cm}
\caption{\label{fig-4}
Total cross section for the process $\gamma^* p \to \eta_C p$
integrated over the region $0<|t|<|t_0|$ as a function of $|t_0|$.
}
\end{figure}


\begin{figure}[htb]
\begin{center}
\epsfig{file=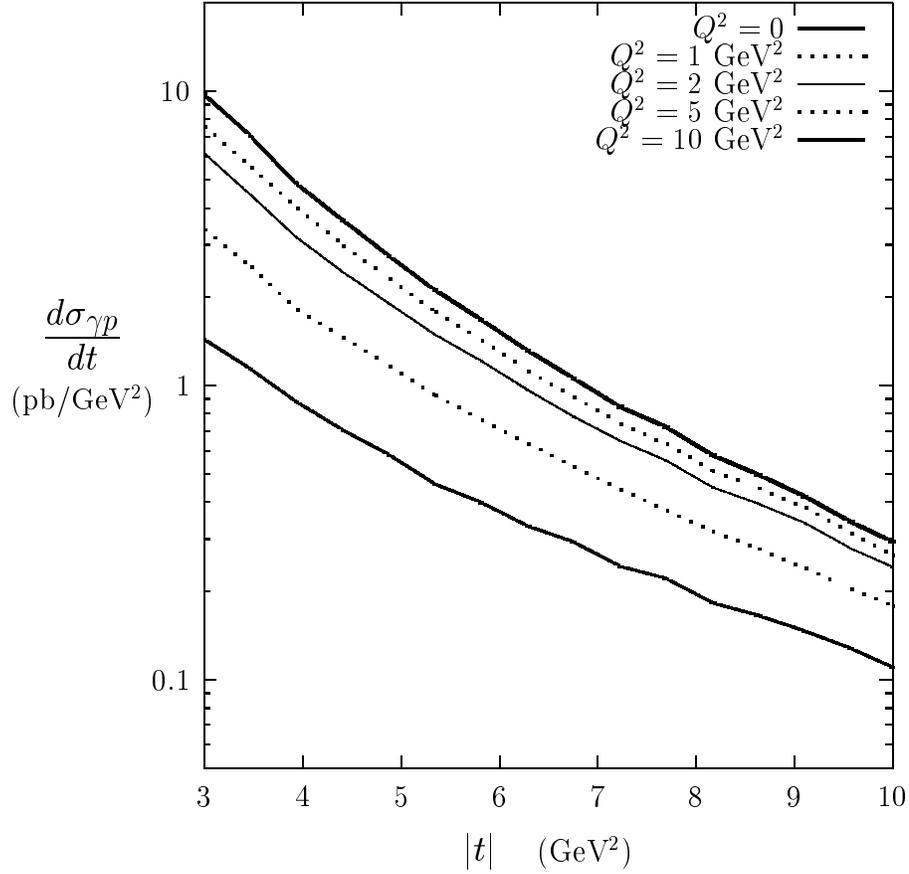,width=12cm}
\end{center}
\vspace*{0.5cm}
\caption{\label{fig-5}
Differential cross section for inelastic process 
$\gamma^* p \to \eta_C X_p$. The strong 
coupling has been evaluated with the scale $\mu^2=|t|$ and 
$\Lambda^{(4)}_{\rm QCD}=0.2 \mbox{ GeV}$.
The integral over the quark densities in the proton has been restricted to 
$x\geq 0.1$.
}
\end{figure}


\begin{figure}[htb]
\begin{center}
\epsfig{file=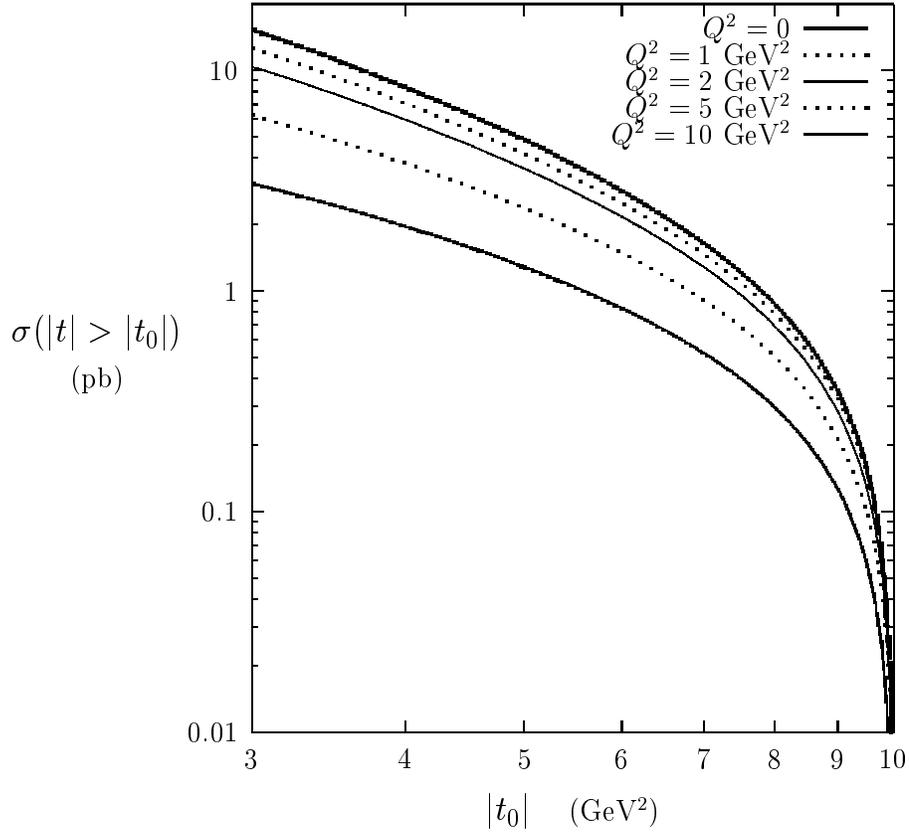,width=12cm}
\end{center}
\vspace*{0.5cm}
\caption{\label{fig-6}
Total cross section for the process $\gamma^* p \to \eta_C X_p$
integrated over the region $|t_0|<|t|$. 
The integral over the quark densities in the proton has been restricted to 
$x\geq 0.1$.
}
\end{figure}


\begin{figure}[htb]
\begin{center}
\epsfig{file=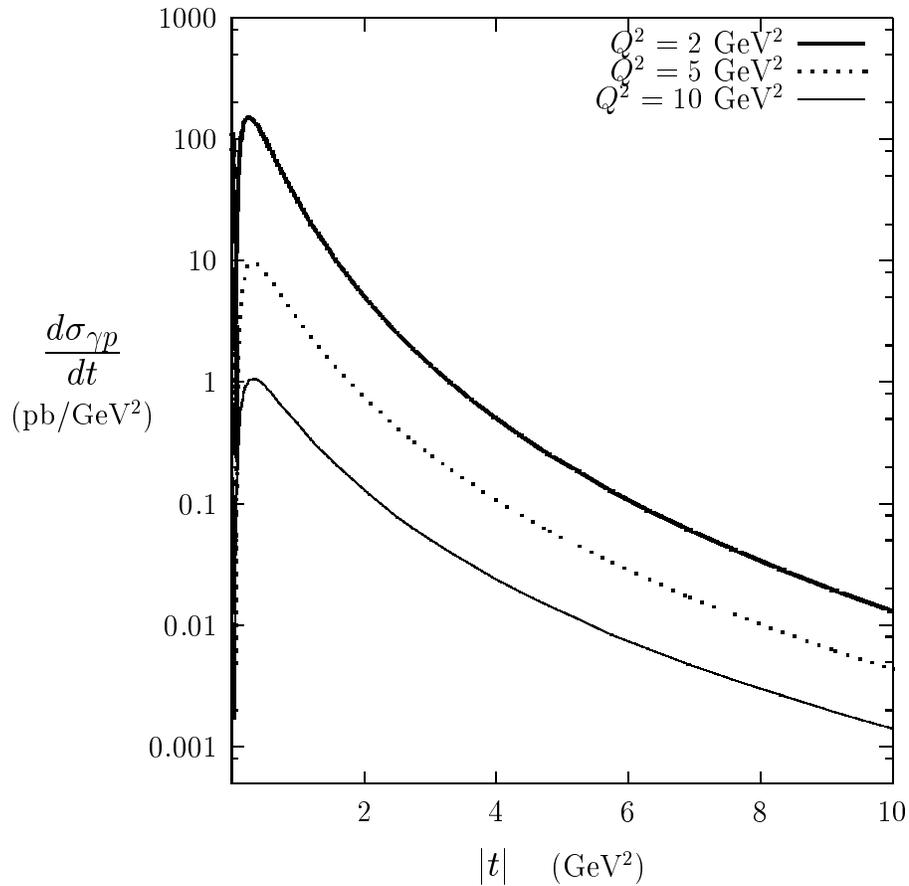,width=12cm}
\end{center}
\vspace*{0.5cm}
\caption{\label{fig-7}
The differential cross section for the 
process $\gamma^* p \to \pi^0 p$. The pion wave function has been approximated 
by the asymptotical form. The strong 
coupling constant has been evaluated with the scale $\mu^2=Q^2 + |t|$ and 
with $\Lambda^{(4)}_{\rm QCD}=0.2 \mbox{ GeV}$.
}
\end{figure}


\begin{figure}[htb]
\begin{center}
\epsfig{file=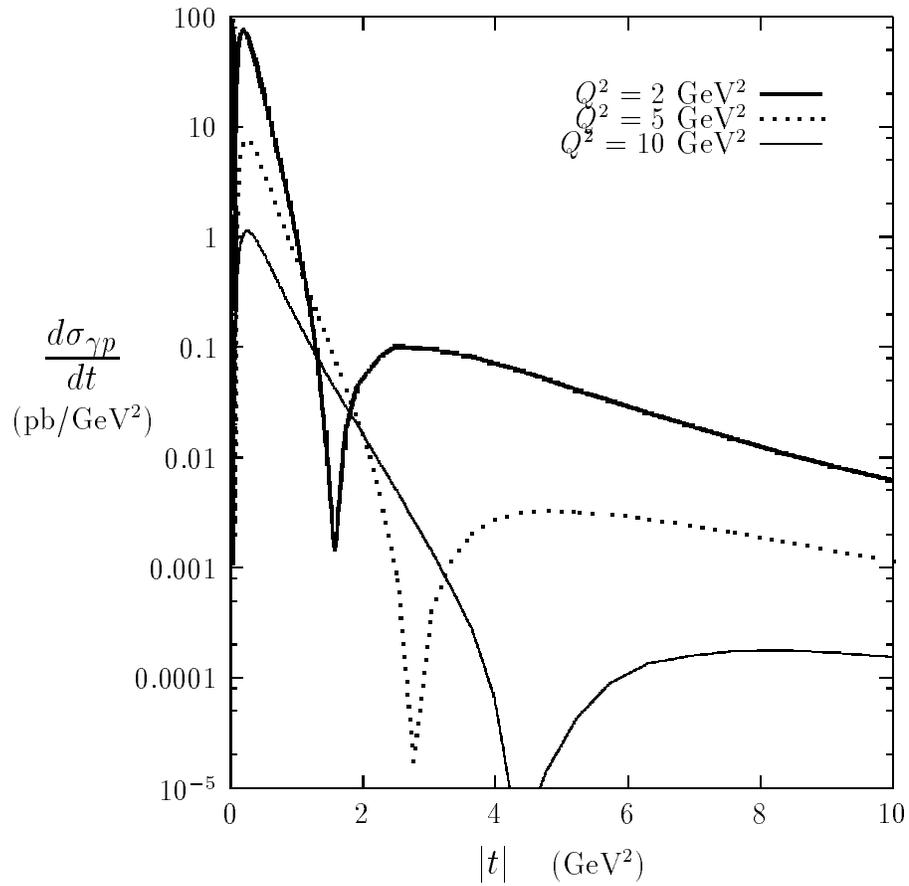,width=12cm}
\end{center}
\vspace*{0.5cm}
\caption{\label{fig-8}
The differential cross section for the 
 process $\gamma^* p \to a_2 p$.  The strong 
coupling has been evaluated with the scale $\mu^2=Q^2 + |t|$ and 
with $\Lambda^{(4)}_{\rm QCD}=0.2 \mbox{ GeV}$.
}
\end{figure}


\end{document}